\documentclass[10.5pt,twocolumn]{article}
\usepackage[left=15mm, top=25mm, bottom=25mm, right=15mm]{geometry}
\usepackage{xcolor}
\usepackage{sectsty}
\usepackage{graphicx} 
\usepackage{amsmath}
\usepackage[superscript,nomove,sort]{cite}
\usepackage{comment}
\usepackage{textcomp}
\usepackage{lineno}
\makeatletter
\renewcommand{\@biblabel}[1]{#1.}
\makeatother

\begin{document}
\title{X-ray pumping of the $^{229}$Th nuclear clock isomer}

\author{
Takahiko  Masuda{$^1$}, 
Akihiro  Yoshimi{$^1$}, 
Akira  Fujieda{$^1$},
Hiroyuki  Fujimoto{$^2$},
Hiromitsu  Haba{$^3$},\\
Hideaki  Hara{$^1$},
Takahiro  Hiraki{$^1$},
Hiroyuki Kaino{$^1$},
Yoshitaka  Kasamatsu{$^4$},
Shinji  Kitao{$^5$}, \\
Kenji  Konashi{$^6$},
Yuki  Miyamoto{$^1$},
Koichi  Okai{$^1$},
Sho  Okubo{$^1$}, 
Noboru  Sasao{$^1$}\thanks{Corresponding author: sasao@okayama-u.ac.jp}, 
Makoto  Seto{$^5$}, \\ 
Thorsten  Schumm{$^7$},
Yudai  Shigekawa{$^4$},
Kenta  Suzuki{$^1$}, 
Simon  Stellmer{$^7$}\thanks{Present address: Physikalisches Institut, Universit\"at Bonn, 53115 Bonn, Germany}, 
Kenji  Tamasaku{$^8$}, \\
Satoshi  Uetake{$^1$},
Makoto  Watanabe{$^6$},
Tsukasa  Watanabe{$^2$},
Yuki  Yasuda{$^4$}, 
Atsushi  Yamaguchi{$^3$},\\
Yoshitaka  Yoda{$^9$},
Takuya Yokokita{$^3$},
Motohiko  Yoshimura{$^1$},
Koji  Yoshimura{$^1$}\thanks{Corresponding author: yosimura@okayama-u.ac.jp}\\[5mm]  
{$^1$}Research Institute for Interdisciplinary Science, Okayama University, Okayama, 700-8530, Japan\\
{$^2$}National Institute of Advanced Industrial Science and Technology (AIST), 1-1-1 Umezono, \\Tsukuba, Ibaraki 305-8563, Japan\\
{$^3$}RIKEN, 2-1 Hirosawa, Wako, Saitama 351-0198\\
{$^4$}Graduate School of Science, Osaka University, Toyonaka, Osaka 560-0043, Japan\\
{$^5$}Institute for Integrated Radiation and Nuclear Science, Kyoto University, Kumatori-cho,\\
      Sennan-gun, Osaka 590-0494, Japan\\
{$^6$}Institute for Materials Research, Tohoku University, Higashiibaraki-gun, Ibaraki 311-1313, Japan\\
{$^7$}Institute for Atomic and Subatomic Physics, TU Wien, 1020 Vienna, Austria\\
{$^8$}RIKEN SPring-8 Center, 1-1-1 Kouto, Sayo-cho, Sayo-gun, Hyogo, 679-5198, Japan\\
{$^9$}Japan Synchrotron Radiation Research Institute, 1-1-1 Kouto, Sayo-cho, \\
Sayo-gun, Hyogo, 679-5198, Japan\\
}
\date{\today}

\twocolumn[
  \begin{@twocolumnfalse}
    \maketitle
    \begin{abstract}
     	Thorium-229 is a unique case in nuclear physics: 
        it presents a metastable first excited state $^{229{\rm m}}$Th, 
	just a few electronvolts above the nuclear ground state.
	This so-called isomer is accessible by VUV lasers, 
	which allows transferring the amazing precision of atomic laser spectroscopy to nuclear physics. 
	Being able to manipulate the $^{229}$Th nuclear states at will opens up a multitude of prospects, 
	from studies of the fundamental interactions in physics to applications as a compact and robust nuclear clock.
	However, direct optical excitation of the isomer or its radiative decay back to the ground state has not yet been observed, 
	and a series of key nuclear structure parameters such as the exact energies and half-lives of 
	the low-lying nuclear levels of $^{229}$Th are yet unknown.
	Here we present the first active optical pumping into $^{229{\rm m}}$Th. 
	Our scheme employs narrow-band 29\,keV synchrotron radiation 
	to resonantly excite the second excited state, which then predominantly decays into the isomer. 
	We determine the resonance energy with 0.07\,eV accuracy, measure a half-life of 82.2\,ps, an excitation linewidth of 1.70\,neV, 
	and extract the branching ratio of the second excited state into the ground and isomeric state respectively. 
	These measurements allow us to re-evaluate gamma spectroscopy data that have been collected over 40~years.
    \end{abstract}
    \vspace{1cm}
  \end{@twocolumnfalse}
]

\section{Introduction}
The first excited nuclear state of $^{229}$Th 
is known to be an isomeric state $^{229\rm m}$Th (metastable excited state).
It has been fascinating the scientific community for decades because 
its energy is expected on the order of only a few eV~\cite{Beck2007,Beck2009, Wense2016},
making it a unique laser-accessible state; in fact it is the lowest nuclear excited state found in Nature so far.
Although the optical excitation from the ground state to $^{229\rm m}$Th is yet to be established experimentally,
aspirations for exploiting it as a new platform for a variety of investigations are expanding.
One important application is an ultra-precise clock.
Such a ``nuclear clock" may reach a fractional uncertainty of $\sim\,10^{-19}$; combined with an increased robustness against perturbations by the outer environments 
it may rival the most precise current optical atomic clocks based on electronic shell transitions~\cite{Peik2003, Kazakov2012, Campbell2012}.
With this precision, it is possible to detect relativistic effects such as a tiny geopotential differences ($\sim$1\,mm)~\cite{Takano2016}; thus the nuclear clock may become a new tool for geodesy. 
In addition, it provides a sensitive probe for investigating the temporal and spatial constancy of fundamental physical
forces such as quantum chromodynamics (QCD) parameters or the electromagnetic coupling~\cite{Flambaum2006, Hayes2007, Berengut2009, Thielking2018}, 
which may be affected by the accelerating universe.

\begin{figure}[tb]
\begin{center}
\includegraphics[width=7.5cm, viewport=0 0 500 400, clip]{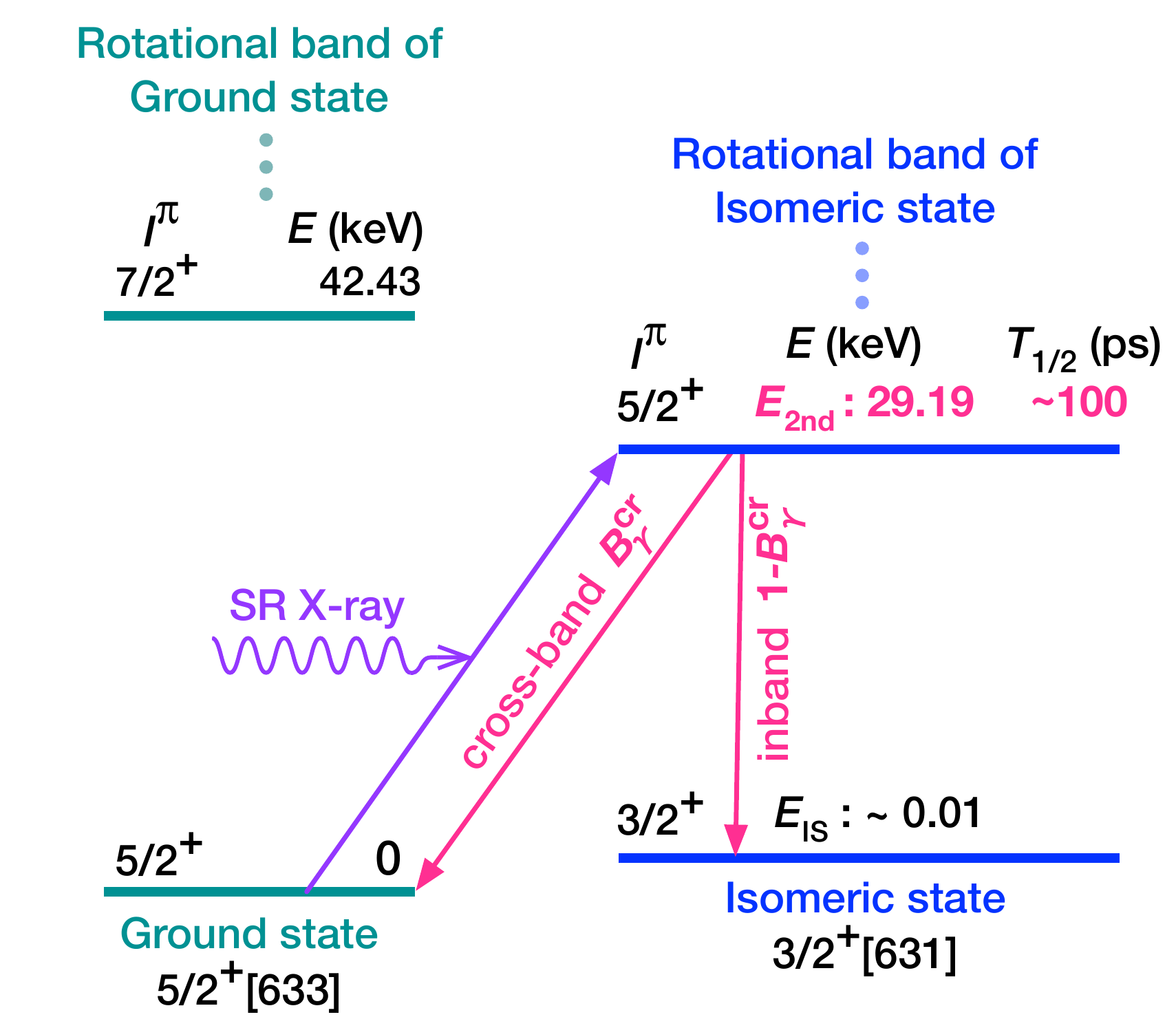}
\caption{Low-energy nuclear level structure of $^{229}$Th. 
Nuclear levels are grouped into two rotational bands, labelled by their band heads; 
$5/2^{+}[633]$ for the ground state and $3/2^{+}[631]$ for the isomeric state.
The $^{229{\rm m}}$Th is populated via X-ray pumping to the second excited state. The quantities in magenta ($E_{\rm 2nd}$, $T_{1/2}$, and $B_{\gamma}^{\rm cr}$) are measured in this work.
\label{fig:energy-level}}
\end{center}
\end{figure}
%

To date, no experiment has unambiguously succeeded in observing decay photons from the isomeric state, 
nor in measuring its energy ($E_{\rm is}$) accurately enough for direct excitation by a narrow band laser.
Recent experiments, however, have gradually constrained the possible $E_{\rm is}$ value.
A $\gamma$-ray spectroscopy measurement using a high-precision calorimeter reported $E_{\rm is}=7.8\pm 0.5$\,eV~\cite{Beck2007,Beck2009}; 
it was obtained as a difference of $\gamma$-ray energies from excited states of $^{229}$Th (including the second excited state).
More recently, an electronic decay channel of the isomeric state was observed 
by detecting electrons produced through an internal conversion process; 
the result indicated $E_{\rm is}$ in the range of $6.3~{\rm eV} < E_{\rm is} < 18.3~{\rm eV}$~\cite{Wense2016}.

All investigations mentioned above utilize $\alpha$-decay of $^{233}$U to produce $^{229{\rm m}}$Th. 
Several experiments attempted a direct optical excitation with broad band $\sim 7.8$\,eV synchrotron radiation~\cite{Jeet2015, Yamaguchi2015, Stellmer2018}; all showed null results, 
suggesting that $E_{\rm is}$ or its half-life might lie outside the commonly expected range. 

In this article, we report direct optical X-ray excitation of $^{229}$Th to the second excited state (29-keV level, see Fig.~\ref{fig:energy-level}).
Beyond resonantly exciting the 29-keV level for the first time, we determine its key nuclear parameters such as the energy $E_{\rm 2nd}$, half-life $ T_{\rm 1/2}$, and excitation linewidth $\Gamma_{\gamma}^{\rm cr}$.
In particular, we have determined $E_{\rm 2nd}$ with an absolute accuracy of 0.07\,eV; the value is important because 
it gives direct access to the isomer energy $E_{\rm is}$ when combined with previous or upcoming $\gamma$-ray measurements.

A large fraction ($\sim$\,58\,\%, see Appendix~\ref{appendix:PHY}) of the 29-keV level population quickly ($\sim$\,100\,ps) decays into the isomeric state.
Thus the scheme realises an active method of populating the isomeric state and enables us 
to measure the isomer's various properties in a well-controlled way.

The paper is organized as follows. 
Following this section, the experimental method is described in Sec.2. 
The analysis and results are presented in Sec.3. Finally, the summary and discussions are given in Sec.4.
There are two appendices: a detailed description of the detector system is
presented in Appendix A, the data taking procedure
and derivation of physics quantities is described in Appendix B.

\section{Experiment}

\begin{figure*}[t]
\begin{center}
\includegraphics[width=17cm, bb=0 0 850 620, clip]{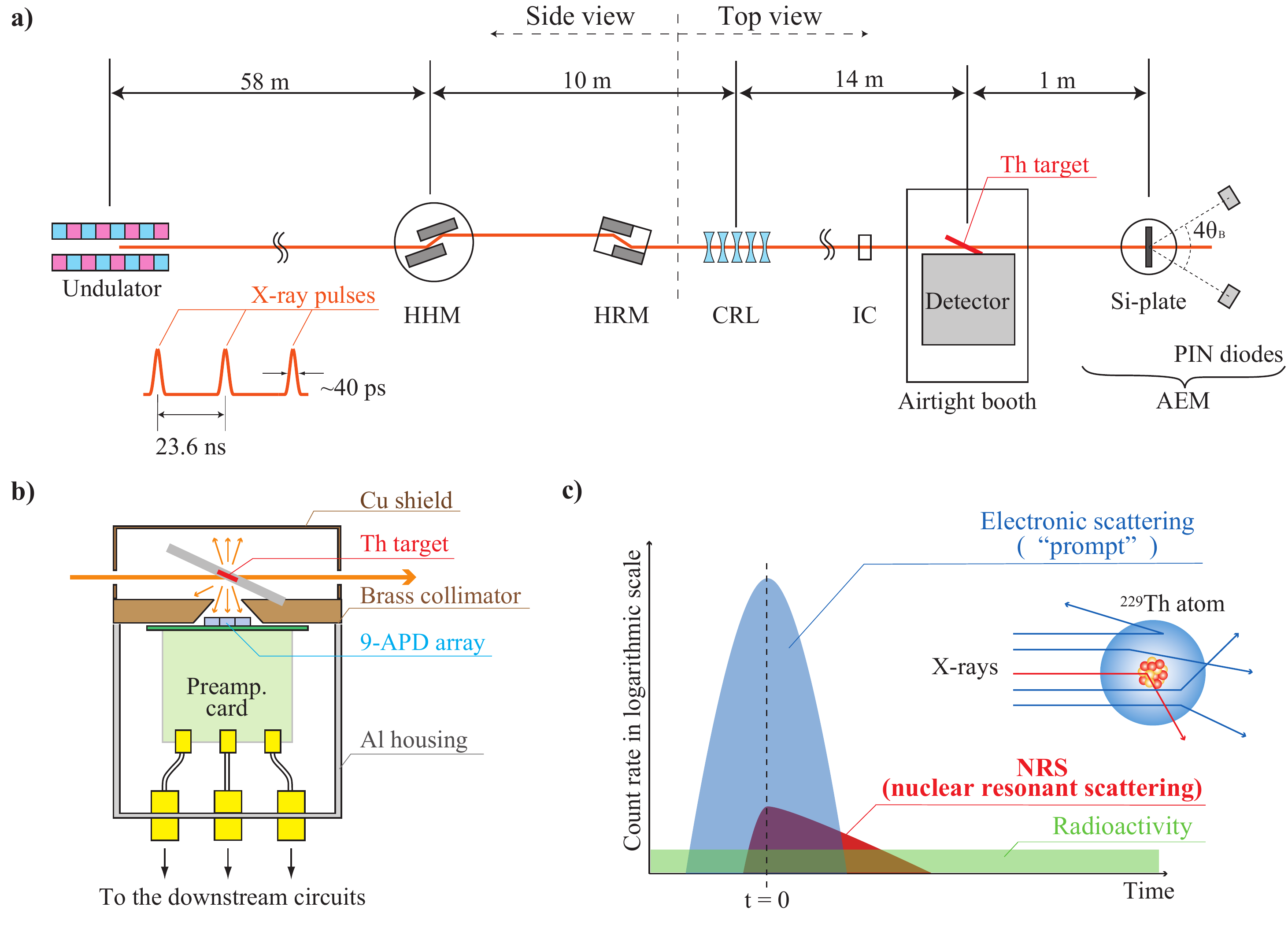}
\caption{Experimental layout and the NRS measurement principle. 
a) Overview of the beam line.  
HHM: High heat-load monochromator, HRM: High resolution monochromator, CRL: Compound refractive lenses, IC: Ionization chamber, AEM: Absolute Energy Monitor using the Bond method.
b) Cut-out view of the $^{229}$Th target and the X-ray detector part.  
c) Illustration of the temporal profiles for the NRS signal (red), the 
  prompt peak (blue), and radioactivity backgrounds (green), together with the corresponding physics processes. 
  The height of the prompt peak is $\sim$6 orders of magnitude larger than the NRS signal.
}
\label{fig:expoverview}
\end{center}
\end{figure*}

\subsection{Measurement principle}
Prior to this work, the energy of the 29-keV level was known to within a few eV only~\cite{NSDF}. 
Using a narrow band X-ray source, the excitation energy must be scanned to search for the nuclear resonance. 
Once the 29-keV level is excited, it decays on the timescale of the half-life, 
either to the isomeric or ground state, respectively, predominantly via an internal conversion process.
Subsequently, various characteristic X-rays are emitted, among which the L-shell lines are detected as the signal. 
An enhancement of this signal indicates the nuclear resonance when the incident X-ray energy is varied.

Conceptually, this scheme is identical to nuclear resonant scattering (NRS)~\cite{Seto2013}.
The present experiment, however, requires several specific developments to account for 
the short half-life ($\sim$\,100\,ps, shortest half-life ever measured in NRS) 
and the extremely small signal-to-background ratio ($\sim 10^{-6}$) due to the narrow excitation linewidth of the 29-keV level. 
A state-of-the-art detector system, specifically developed for this experiment, 
together with an enhanced luminosity realised by a small-spot-size beam and target, 
were the key to success, as will be elaborated below.

\subsection{Experimental setup}
The experiment was carried out at the BL19LXU beam line of SPring-8~\cite{BL19LXU}. 
Schematic drawings of the beam line and setup are shown in Fig.~\ref{fig:expoverview}.

The high-brilliance X-ray beam line starts with a 27-m-long undulator.
The produced X-ray photons are monochromatised with two pairs of Si crystals, Si(111) and Si(440), inserted in series into the beam line.
The intensity and full-width-at-half-maximum (FWHM) bandwidth after Si(440) is approximately $4\times 10^{12}$\,photons/s and 0.26\,eV, respectively.
After identifying the resonance, the second monochromator is replaced by Si(660) which reduces the bandwidth to 0.10\,eV.
A lens system, Compound Refractive X-ray Lens (CRL)~\cite{CRL}, is employed to focus the beam to a spot size of 0.15\,$\times$\,0.065\,mm$^2$ at the focal point, $\sim$\,14\,m downstream from the device, 
with a transmission of about 67\,\%.
The actual X-ray beam is a pulse train, with pulses separated by 23.6\,ns and 40\,ps pulse duration.
Typical properties of the monochromators, along with the $^{229}$Th target, are summarized in Table~\ref{tab:beamsummary}.

\renewcommand{\arraystretch}{1.2}
\begin{table}[h]
\begin{center}
\caption{Properties of the 29 keV X-ray beam and $^{229}$Th target.}\label{tab:beamsummary}
\begin{tabular}{p{44mm}p{9mm}p{9mm}p{9mm}}
\hline
Monochromator & Si(111) & Si(440)  & Si(660) \\[1mm]
\hline
Intensity ($10^{12}$ photons/sec)& \multicolumn{1}{c}{$80$} & \multicolumn{1}{c}{$4$} & \multicolumn{1}{c}{$1$} \\[1mm]
Energy bandwidth (eV) & \multicolumn{1}{c}{3.4}  & \multicolumn{1}{c}{0.26}  & \multicolumn{1}{c}{0.10}  \\
Beam size w/o lens (mm)& \multicolumn{3}{c}{1.5 (h) $\times$ 0.8 (v)}\\[1mm]
Beam size with lens (mm) & \multicolumn{3}{c}{0.15 (h) $\times$ 0.065 (v)}\\[1mm]\hline 
Target size (mm) & \multicolumn{3}{c}{0.2 (t) $\times$ 0.4 ($\phi$)}\\[1mm]
Target total amount ($\mu$g) & \multicolumn{3}{c}{0.24}\\[1mm]
\hline
\end{tabular}
\end{center}
\end{table}


The $^{229}$Th target with a small diameter ($\phi$\,=\,0.4\,mm)  is prepared by a dry-up method; 
$^{229}$Th solution (0.1\,mol/L HNO$_3$) is poured  into a groove machined into a thin graphite plate and dried up by heating.
A total of 0.24\,$\mu$g of $^{229}$Th (1.8\,kBq, 6.3\,$\times 10^{14}$\,nuclei) is deposited and hermetically sealed with Be cover plates.
The target is placed at the focal point of the beam at an angle of 22$^{\circ}$, 
as shown in Fig.~\ref{fig:expoverview}b.

The X-ray fluorescence emitted by the target after exposure is detected by silicon avalanche photodiode (Si-APD) sensors (see Fig.~\ref{fig:expoverview}b). 
The detector, consisting of 9 APD chips ($\phi$\,=\,0.5\,mm, S12053-05, Hamamatsu Photonics) arrayed in a $3\times3$ matrix, 
is placed at a distance of 3.5\,mm from the target centre, covering 0.95\,\% of the solid angle.
The output signals are amplified by on-board preamplifiers and then sent to subsequent signal processing circuits, 
where energy and arrival time of each photon are recorded.
The resolutions on these quantities are $\Delta E/E\simeq 21\%$~\cite{Masuda2019} (FWHM)
 and $\Delta t\simeq$120\,ps~\cite{Masuda2017}(FWHM) at 13\,keV, the energy of 
 the dominant $^{229}$Th characteristic X-ray line relevant for detecting the NRS signals 
 (see Appendix~\ref{appendix:APD} for more details).

The energy of the incident beam is regularly monitored during the measurements 
by an absolute energy monitor (AEM) placed downstream of the $^{229}$Th target (see Fig.~\ref{fig:expoverview}a). 
This device measures a pair of the Bragg angles, formed on the left and right sides of the incident beam,
 with a Si(440) reference crystal mounted on a high-precision rotary table. 
With this method, developed originally by Bond~\cite{Bond1960}, the energy can be determined with an accuracy of 0.07\,eV 
(see Appendix~\ref{appendix:AEM} for more details).

\subsection{Properties of signal and backgrounds \label{sec:Properties of signal}}
In order to isolate the NRS signal from the overwhelming background, it is essential to understand its properties, 
in particular its energy spectrum and temporal behaviour. 
Figure~\ref{fig:expoverview}c illustrates a temporal profile of the NRS signal and various backgrounds 
along with the corresponding physical processes.

The decay of the 29-keV level occurs predominantly through the internal conversion process followed by emission of characteristic X-rays; 
the dominant components of these X-rays are the $L_{\alpha}$ and $L_{\beta}$ lines whose energy are, respectively, 
13\,keV and 16\,keV \cite{Raboud1999,IsotopeTable} (see Appendix~\ref{appendix:PHY} for more details). 
When the incident energy coincides with the resonance energy, 
the NRS signal appears in the form of an exponential slope in the temporal profile, reflecting the half-life of the excited state. 
In contrast, most background events occur at $t=0$, the time each incident X-ray pulse passes through the target 
(thus referred to as ``prompt"). 
These events are mainly due to the 
photoelectric absorption process occurring in the $^{229}$Th electronic shell, followed by X-ray fluorescences. 
In addition, there is a contribution to the prompt peak from other processes such as Rayleigh and/or Compton scattering. 
See the inset of Fig.~\ref{fig:lifetime}\, for a measured prompt peak.

Although weaker, other types of backgrounds, having different timing properties, also exist: 
one is caused by the radioactivity of $^{229}$Th and 
its daughter nuclei (random in time) and another by stray photons scattered off surrounding materials 
(definite time, depending on the distance to the material). 
The energy of these backgrounds are, in general, different from that of the signal 
except for the photoelectric absorption process which has an energy spectrum similar to that of the signal 
(see Extended Data Fig.~\ref{fig:Energy-Spectrum-prompt-NRS}).
Note that the prompt signal exceeds the NRS signal by 6 orders of magnitudes.

\section{Analysis and results}
The data presented below was taken in July (Run 1) and November (Run 2) 2018.
Initially the Si(440) monochromator was used to search for the 29-keV level.
After identifying its resonance energy, Si(440) was replaced by Si(660) which has a smaller bandwidth (see Table~\ref{tab:beamsummary}).
All physics quantities such as the energy, half-life, excitation linewidth of the 29-keV level, are  
derived from the Si(660) data unless stated otherwise.   
For more details on the data taking and evaluation procedure, see Appendices~\ref{appendix:DAT} and \ref{appendix:PHY}.

\subsection{Resonance energy}
\begin{figure}[t]
\begin{center}
\includegraphics[width=9cm, bb=0 0 560 500, clip]{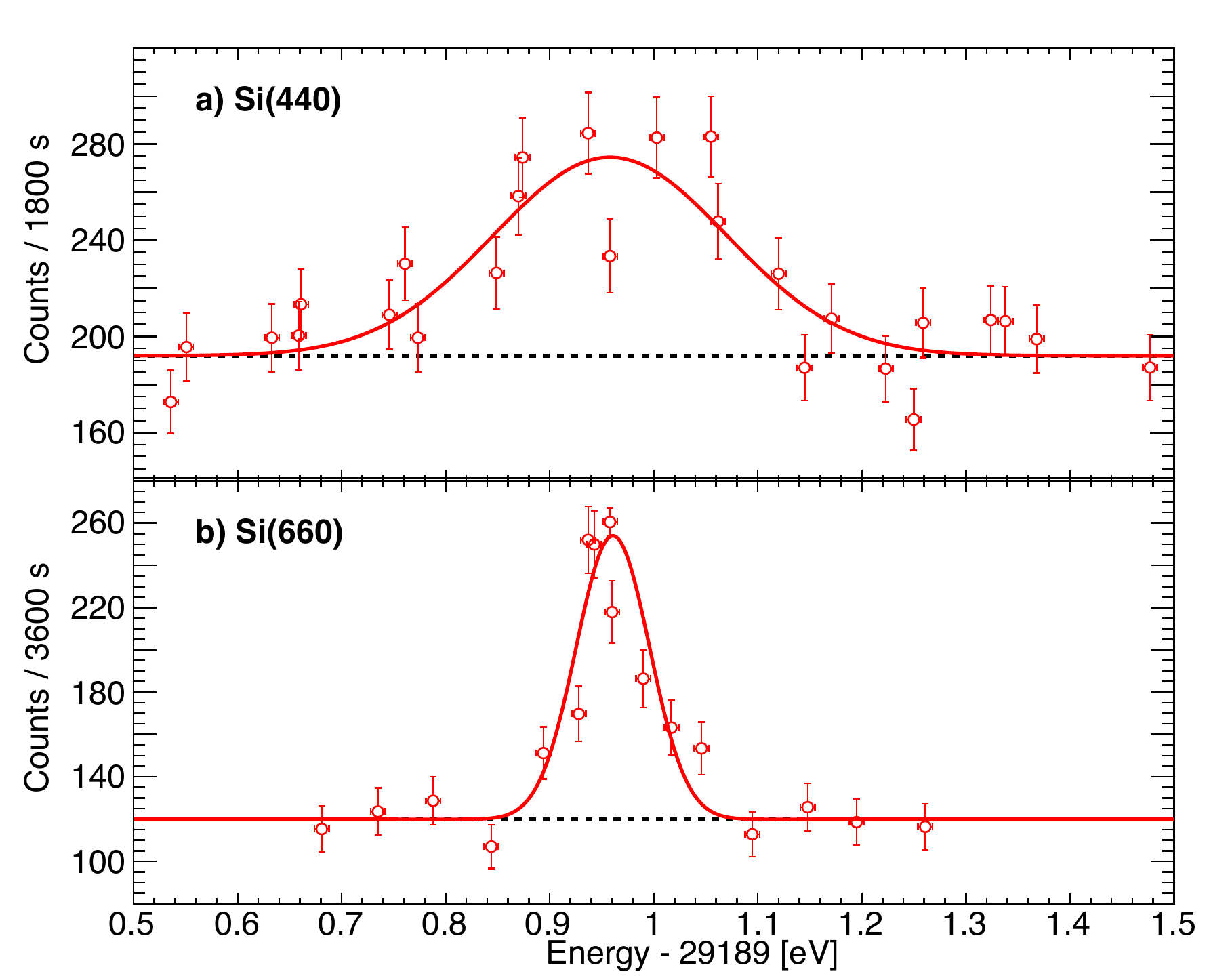}
\caption{Resonance curves of $^{229}$Th.
NRS signals obtained with the Si(440) (a) and Si(660) (b) monochromators. 
The vertical axes indicate the number of events inside the signal window while the horizontal axis depicts the absolute X-ray beam energy 
(offset by 29189\,eV).
The error bars represent the statistical uncertainty. 
The solid curves show the results of a Gaussian fit with a constant background (dashed lines). 
Each data point corresponds to one standard run (1800\,s and 3600\,s, respectively) except for the point at the peak in the lower plot (21600\,s), 
used for the decay measurement in Fig.~\ref{fig:lifetime}. 
}
\label{fig:resonance}
\end{center}
\end{figure}

Successful excitation to the 29-keV level is signalled by an enhancement in the number of 
APD signals within a specified energy-time window (``NRS signal window").
The actual window is set to 12--18\, keV in energy and 
   0.40--0.90\,ns in time after the prompt signal, considering the character of the signal and backgrounds.   
The resonance curve obtained with the Si(440) monochromator is shown in Fig.~\ref{fig:resonance}a. 
A clear NRS peak is observed on a constant background. 
Figure~\ref{fig:resonance}b shows the resonance curve taken with Si(660) in Run 1.
The resonance energy $E_{\rm 2nd}$ is obtained by fitting a Gaussian function plus a constant to the Si(660) data. 
We obtained $29189.961\pm 0.006$\,eV (Run 1) and $29189.908 \pm 0.005$\,eV (Run 2) 
where the indicated uncertainties are statistical.
Taking the weighted average and including all systematic uncertainties, 
the final value is determined as 
\begin{equation}
 E_{\rm 2nd}=29189.93 \pm 0.07 \hspace{1.5mm} \mbox{eV},
\label{eq:E2nd}
\end{equation}
where the error is dominated by the uncertainty in determining the absolute energy of the incident beam (see Appendix~\ref{appendix:AEM}).

The width of the resonance curve is fully determined by the monochromator bandwidth; 
the actual root-mean-square width (rms width, $\sigma_{\rm Xray}$) is found to be $0.112\pm 0.014$  eV for Si(440), 
and $0.041 \pm 0.006$  eV for Si(660), respectively.
Note that in Table~\ref{tab:beamsummary} the FWHM bandwidth is listed, which is $2\sqrt{2 \log 2}\; \sigma_{\rm Xray}$.

\subsection{Half-life}
\begin{figure}[t]
\begin{center}
\includegraphics[width=9cm, bb=0 0 550 450, clip]{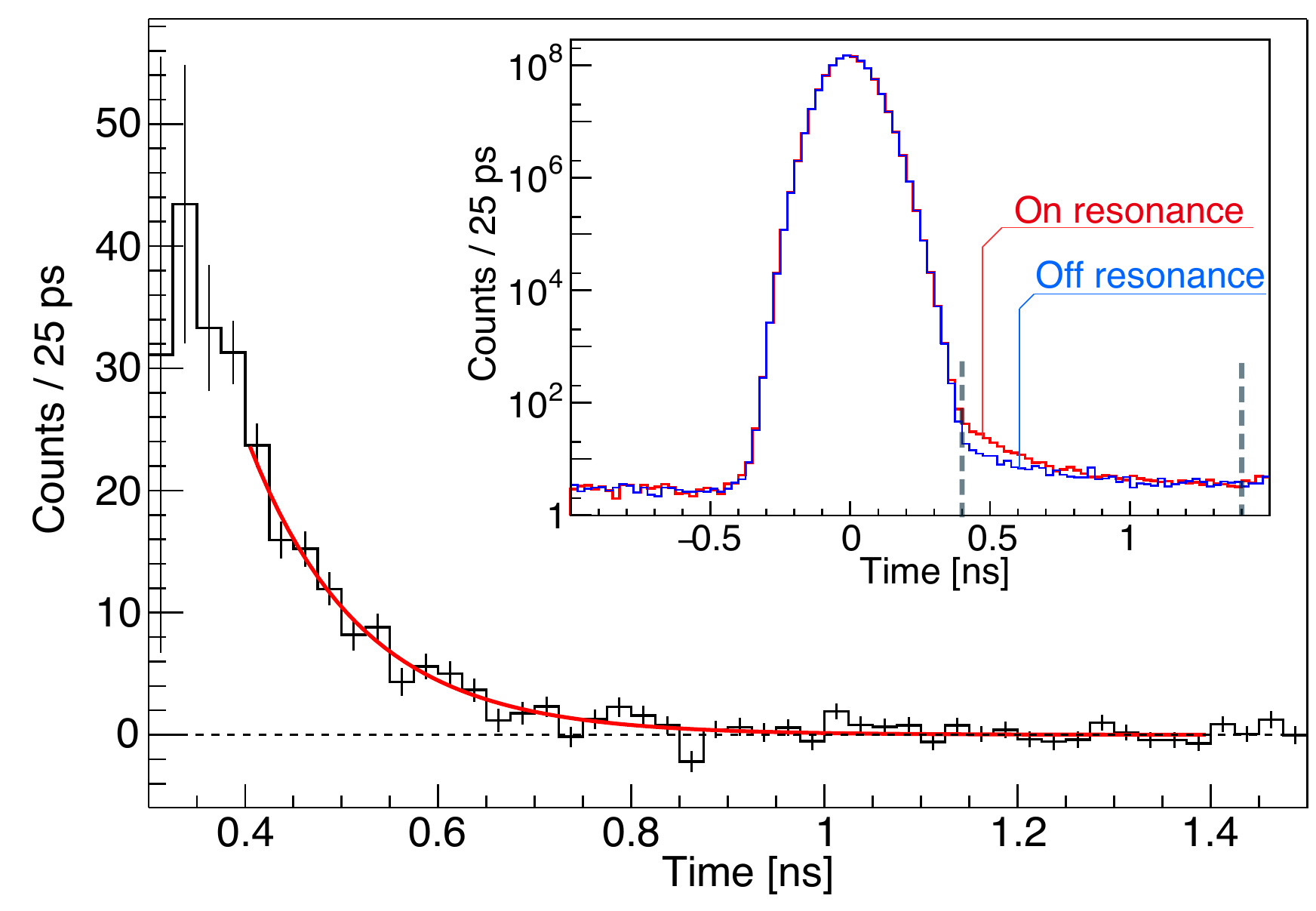}
\caption{Lifetime of the $^{229}$Th second excited state.
(inset) NRS signal temporal profiles of the Si(660) data on resonance (red) and off resonance (blue), 
normalized to a 3600\,s run. 
(lower plot) The subtracted signal in an expanded time region, error bars represent the statistical uncertainty. 
The red solid curve shows the result of a simple exponential fit to the region from 0.4 to 1.4\,ns (indicated by vertical dashed lines in the inset).
}
\label{fig:lifetime}
\end{center}
\end{figure}

The half-life of the second excited state of the $^{229}$Th nucleus is obtained 
with on- and off-resonance data taken with the Si(660) monochromator.
The inset of Fig.~\ref{fig:lifetime} shows the on- (red) 
and off-resonance (blue) temporal profiles of Run 1.
The NRS time signal is obtained by subtracting the off-resonance data from on-resonance data. 
As seen in Fig.~\ref{fig:lifetime}, the signal exhibits a clear exponential decay. 
An exponential fit to the region from 0.4 to 1.4\,ns yields a half-life of $84.2 \pm 6.3$\,ps 
(red solid curve in Fig.~\ref{fig:lifetime}).
Similarly, the Si(660) data from Run 2 yields $80.9 \pm 5.1$\,ps. 
Taking the weighted average of these results, the half-life is 
\begin{eqnarray}
 T_{1/2}=82.2 \pm 4.0 \hspace{1.5mm}{\rm ps}.
 \label{eq:half life time} 
\end{eqnarray}
The present result significantly supersedes the shortest half-life hereto measured in NRS spectroscopy
 (it was $\sim 630$\,ps in $^{201}$Hg)~\cite{Ishikawa2005,Yoshimi2018}, and offers new potential to this field.

\subsection{Excitation linewidth}
Another key quantity of the 29-keV level is the excitation linewidth, {\it i.e.} the radiative transition width between the ground 
and the second excited state (cross-band transition, $\Gamma_{\gamma}^{\rm cr}$). 
This quantity may in principle be derived from the obtained NRS counting rates and absolute 
knowledge of the experimental parameters such as the beam intensity, target density, and APD detection efficiency.
In reality, however, reliable values for all of these parameters are difficult to obtain.
Fortunately, $\Gamma_{\gamma}^{\rm cr}$ can be expressed by the ratio of the NRS and the prompt peak rates, times the well-known photoelectric absorption 
cross section~\cite{NIST-XCOM}. In this ratio, several experimental parameters drop out. 
The actual calculation requires various corrections; the procedure is detailed in Appendix~\ref{appendix:PHY}.
The final result is 
\begin{eqnarray}
 \Gamma_{\gamma}^{\rm cr}=1.70 \pm 0.40\hspace{1.5mm} {\rm neV}.
 \label{eq:radiative width} 
\end{eqnarray}
The obtained width $\Gamma_{\gamma}^{\rm cr}$ can be used to determine the radiative branching ratio 
of the 29-keV level into the isomer and ground state respectively (see Fig~\ref{fig:energy-level}).
To this end, knowledge of the radiative transition width between the 29-keV level and the isomeric state 
(inband transition, $\Gamma_{\gamma}^{\rm in}$) is necessary, which has been determined experimentally in Refs.\cite{Barci2003, Kroger1976} to $\Gamma_{\gamma}^{\rm in}=14.3\pm 1.4$\,neV (averaged value). 
Thus the radiative branching ratio to the ground state is
\begin{eqnarray}
 B_{\gamma}^{\rm cr}=\frac{\Gamma_{\gamma}^{\rm cr}}{\Gamma_{\gamma}^{\rm cr}+\Gamma_{\gamma}^{\rm in}}=\frac{1}{(9.4\pm 2.4)}.
 \label{eq:branching ratio} 
\end{eqnarray}
This value is compatible with 1/13 (with an 8\% error) quoted by Ref.\cite{Beck2007}, 
and lies in the range suggested by Ref.\cite{Tkalya2015}.

\section{Discussion and Summary}

In this work, the energy of the second excited state $E_{\rm 2nd}$ and 
the radiative branching ratio $B_{\gamma}^{\rm cr}$ are determined.
These values can be used to constrain the isomer energy $E_{\rm is}$ when combined with published (or future) $\gamma$-ray spectroscopy data such as Refs.~\cite{Helmer1994, Barci2003, Beck2007, Beck2009}. Reported $\gamma$-measurements of the 29-keV level contain both, a (strong) inband and a (weak) cross-band contribution (29-keV $\gamma$ doublet) which so-far cannot be resolved. 
The measured $\gamma$-energy is hence the weighted sum of the doublet contributions, 
$E_{\gamma}^{\rm dblt}=E_{\rm 2nd}B_{\gamma}^{\rm cr}+(E_{\rm 2nd}-E_{\rm is})(1-B_{\gamma}^{\rm cr})$. 
Knowing $E_{\rm 2nd}$ and $B_{\gamma}^{\rm cr}$, $E_{\rm is}$ can be extracted.

\begin{figure}[tb]
\begin{center}
\includegraphics[width=9.0cm,bb=70 290 525 560, clip]{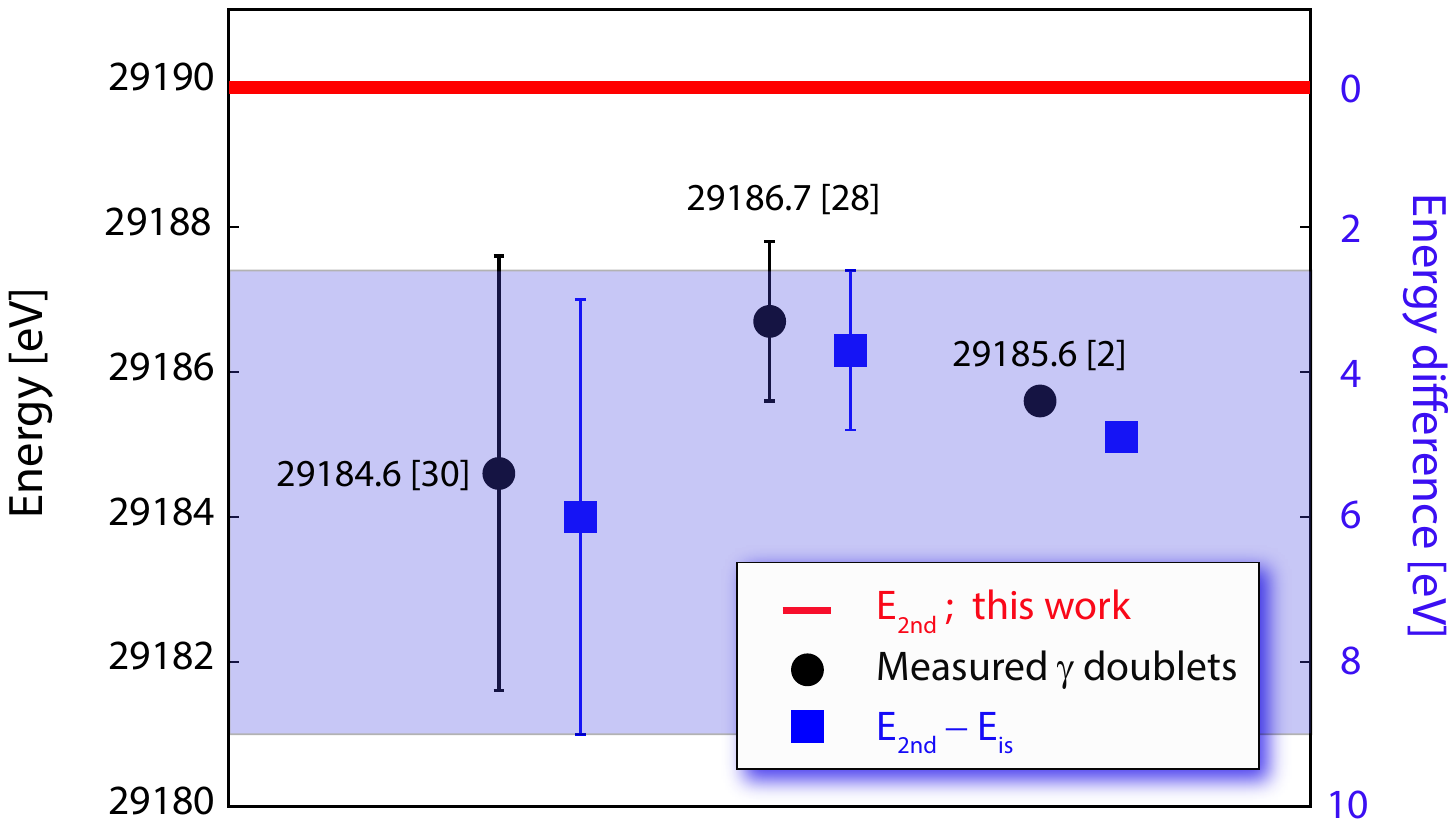}
\caption{Constraining the isomer energy $E_{\rm is}$.
The red bar indicates  $E_{\rm 2nd}$ measured in this work while 
the black circles are results of 29-keV $\gamma$-doublet measurements by 
Ref.~\cite{Helmer1994}, Ref.~\cite{Barci2003} and Ref.~\cite{Beck2007, Beck2009},
from left to right, respectively.
The blue squares represent $E_{\rm 2nd}-E_{\rm is}$, extracted from $E_{\gamma}^{\rm dblt}$ by taking out a weak $\gamma$-signature from cross-band decays, 
connecting the 29-keV level to the ground state.
No error is provided in Ref.~\cite{Beck2007, Beck2009}.
}
\label{fig:energy-comparison}
\end{center}
\end{figure}

Fig.~\ref{fig:energy-comparison} shows our measurement of $E_{\rm 2nd}$ with error ($\pm 0.07$\,eV) indicated by a red horizontal bar.
Experimental values of $E_{\gamma}^{\rm dblt}$ from  $\gamma$-ray spectroscopy measurements are depicted by the black circles with their reported errors.
The  blue squares show the extracted values, taking out the weak cross-band contribution, yielding $E_{\rm 2nd}-E_{\rm is}$, 
where the error bars include the uncertainties in $E_{\gamma}^{\rm dblt}$ and $B_{\gamma}^{\rm cr}$.
The difference from $E_{\rm 2nd}$  (red horizontal bar) to these extracted data points (blue squares) yields $E_{\rm is}$.
Taking the lower and upper limits of the error bars $E_{\rm is}$ is found within the range 
of $2.5\ {\rm eV} < E_{\rm is} < 8.9\ {\rm eV}$.

The result suggests that 
$E_{\rm is}$ is indeed in a laser-accessible energy region.
Any future improvement in the accuracy of $\gamma$-measurements of the 29-keV level will directly lead to a more accurate determination of $E_{\rm is}$. 
Several groups are currently developing dedicated calorimeters for this purpose~\cite{Kazakov2013}.

%

The present pumping scheme realises an efficient method for $^{229\rm m}$Th production.
It has several advantages compared to the scheme using the $^{233}$U $\alpha$-decay chain.

First, the $^{229\rm m}$Th production rate is high; it amounts to 
 25\,kHz in the present experiment (see Appendix~\ref{appendix:PHY} for details).
Obtaining this rate with $^{233}$U $\alpha$-decays requires an activity about 700 times larger 
than that used in our target.

Another advantage is experimental control. 
In $^{233}$U decay $\sim$5\,MeV $\alpha$-energy is released, generating large stochastic background. 84\,keV recoil energy is transferred to the nucleus, leaving it in a largely uncontrolled state concerning its kinetics and ionization level. 
These are challenging conditions for a direct optical excitation or detection of the isomeric state, 
let alone the construction of a nuclear clock. 
So-far only one group worldwide has succeeded in producing a controlled ion beam of $^{229\rm m}$Th~\cite{Wense2016,Seiferle2017	}.

In contrast, optical X-ray pumping via the second excited state transfers a negligible recoil energy of 1.8\,meV to the nucleus, not affecting the charge or motional state.
It is hence compatible with  $^{229\rm m}$Th production within optically transparent samples.
Any isomer-related signals can be unambiguously identified by switching the excitation on and off. 
Furthermore, when $^{229}$Th targets are prepared using ionic states ($^{229}$Th$^{n+}$ with $n\ge2$), the half-life of the isomeric state is expected to be long, so that backgrounds caused by the incident X-ray beam can be almost entirely eliminated.

Direct optical detection and precision spectroscopy of the $^{229}$Th isomer is the next important step towards the realization of the nuclear clock.
Using the presented pumping scheme, we are currently preparing an experiment to detect the vacuum ultra violet (VUV) transition using a $^{229}$Th-doped VUV transparent crystal. 

In summary, we have realised the first active excitation of $^{229\rm m}$Th using an X-ray pumping scheme.
We measured the second excited state energy level $E_{\rm 2nd}$ with an accuracy of 0.07\,eV, and constrained the isomer energy $E_{\rm is}$.
This scheme paves the way for future in-depth investigations of the $^{229}$Th isomeric transitions in a well-controlled way.


\section*{Acknowledgement}
The synchrotron radiation experiments were performed at the BL09XU and the BL19LXU line of SPring-8 
with the approval of the Japan Synchrotron Radiation Research Institute (JASRI) 
(proposals No. 2016B1232, 2017B1335, 2018A1326, and 2018B1436) and RIKEN (No. 20180045).
The authors would like to thank all members of the SPring-8 operation and supporting teams.
The experiment received support from KEK-PF (No. 2017G085) and  
IMR-Tohoku U. (No. 18F0014), 
where indispensable detector tests and target preparation were performed. 
Special thanks should go to Prof. S. Kishimoto for his support at KEK,  
to Mr. T. Kobayashi for technical assistance at SPring-8, and 
to Mr. K. Beeks for valuable discussion during the preparation of the manuscript.
This work was supported by JSPS KAKENHI Grant Numbers JP15H03661, JP17K14291, JP18H01230, and JP18H04353. 
T.S. and S.S.  gratefully acknowledge funding by the EU FET-Open project, Grant No. 664732 “nuClock”.
Ak.Y. and At.Y. acknowledge the MATSUO foundation and Technology Pioneering Projects in RIKEN, respectively.
\vspace{1cm}

\section*{Appendices}
\begin{appendix}
\section{Detector system\label{appendix:Detector}}

\subsection{APD detector and readout system\label{appendix:APD}}
The Si-APD detector is developed in cooperation with Hamamatsu Photonics K.K. based on our custom design. 
It consists of 9 APD chips and is arrayed in a 3$\times$3 matrix of 1.14\,mm pitch between adjacent chips.
With a 0.5-mm-diameter photocathode, placed 3.5\,mm away from the target, the total geometrical acceptance amounts to 0.95\%.

\begin{figure}[t]
\begin{center}
\includegraphics[width=7.5cm, bb=0.0 0.0 344.25 350, clip]{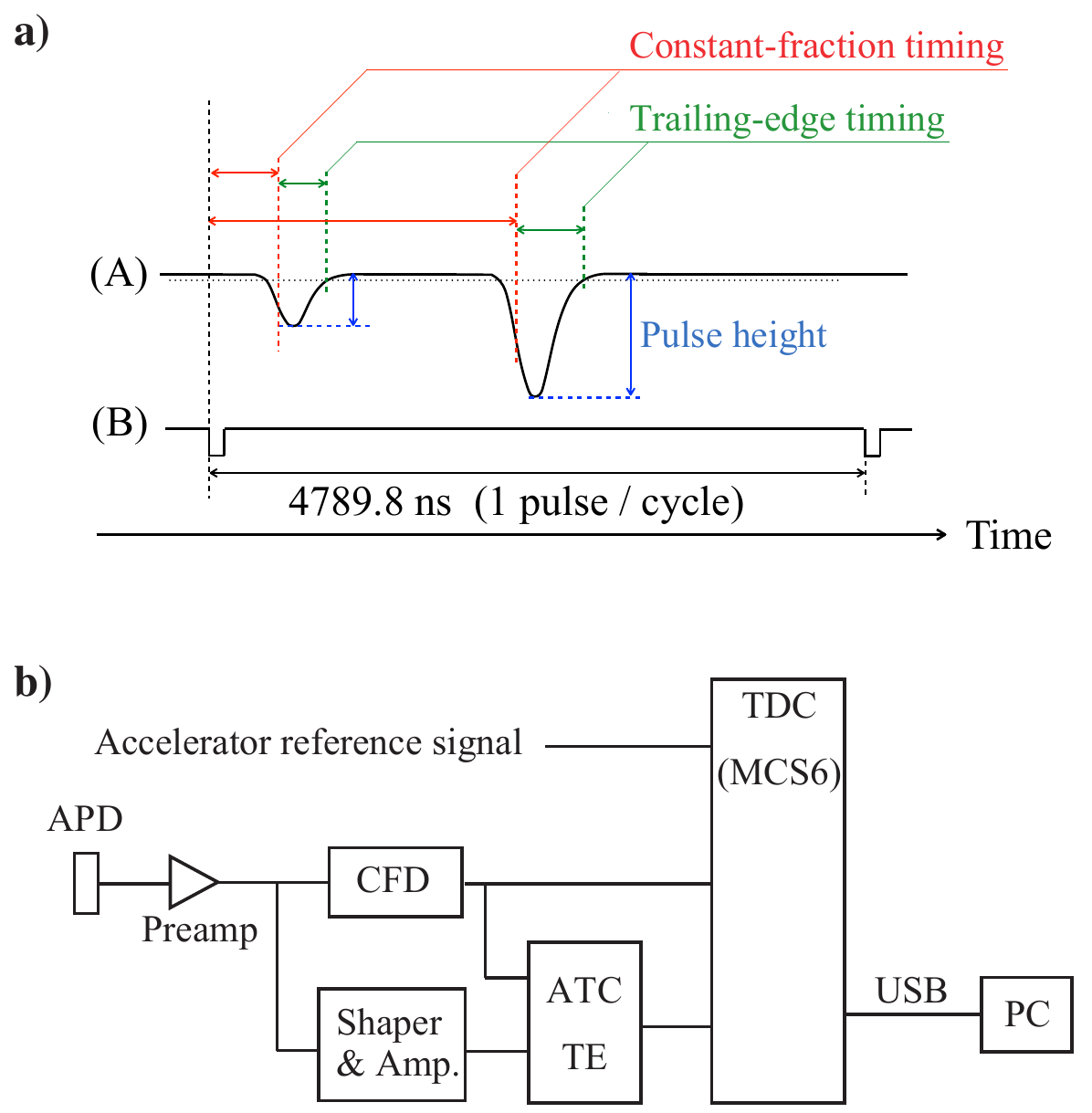}
\caption{Conceptual diagram of the pulse processing scheme.
a) Timing chart.
Line (A) shows analogue pulses from an APD chip and line (B) shows the accelerator reference clock. 
The example shows two pulses with different photon energies within a cycle. 
b) Block diagram.  
For each pulse, three parameters are stored for the post analysis: constant-fraction timing (CFD), 
pulse height (ATC), and trailing-edge timing (TE).
\label{fig:chart}}
\end{center}
\end{figure}

The conceptual diagram of the signal processing is shown in Fig~\ref{fig:chart}. 
Each Si-APD output is amplified by a fast amplifier (Mini-circuits, RAM-8A+) located directly behind the Si-APDs. 
The amplified output is processed in several ways to obtain event characteristics~\cite{Masuda2017, Masuda2019}.
First of all, it is converted to a logic signal by a lab-built constant fraction discriminator (CFD); 
this yields an event  occurrence time.
Second, the signal is converted to a logic signal whose delay time is proportional to the APD output amplitude (ATC); it provides an event energy.
Third, the trailing edge (TE) timing of the analogue signal is also obtained; this is used to reject
events with abnormal waveforms caused by, for example, multiple hits.
All the logic signals (CFD, ATC, TE) are sent to multiple-event time digitizers (MCS6; FAST ComTec GmbH) 
with a time-bin-width of 100\,ps, together with the accelerator clock. 
The clock provides a reference timing of the electron bunch revolution in the accelerator ring (every 4789.8\,ns).
Finally, the digitized signals are all stored in PCs for off-line analysis.

\subsection{Absolute X-ray energy monitor\label{appendix:AEM}}
\begin{figure}[t]
\begin{center}
\includegraphics[width=9cm,  bb=0.0 0.0 950 650, clip]{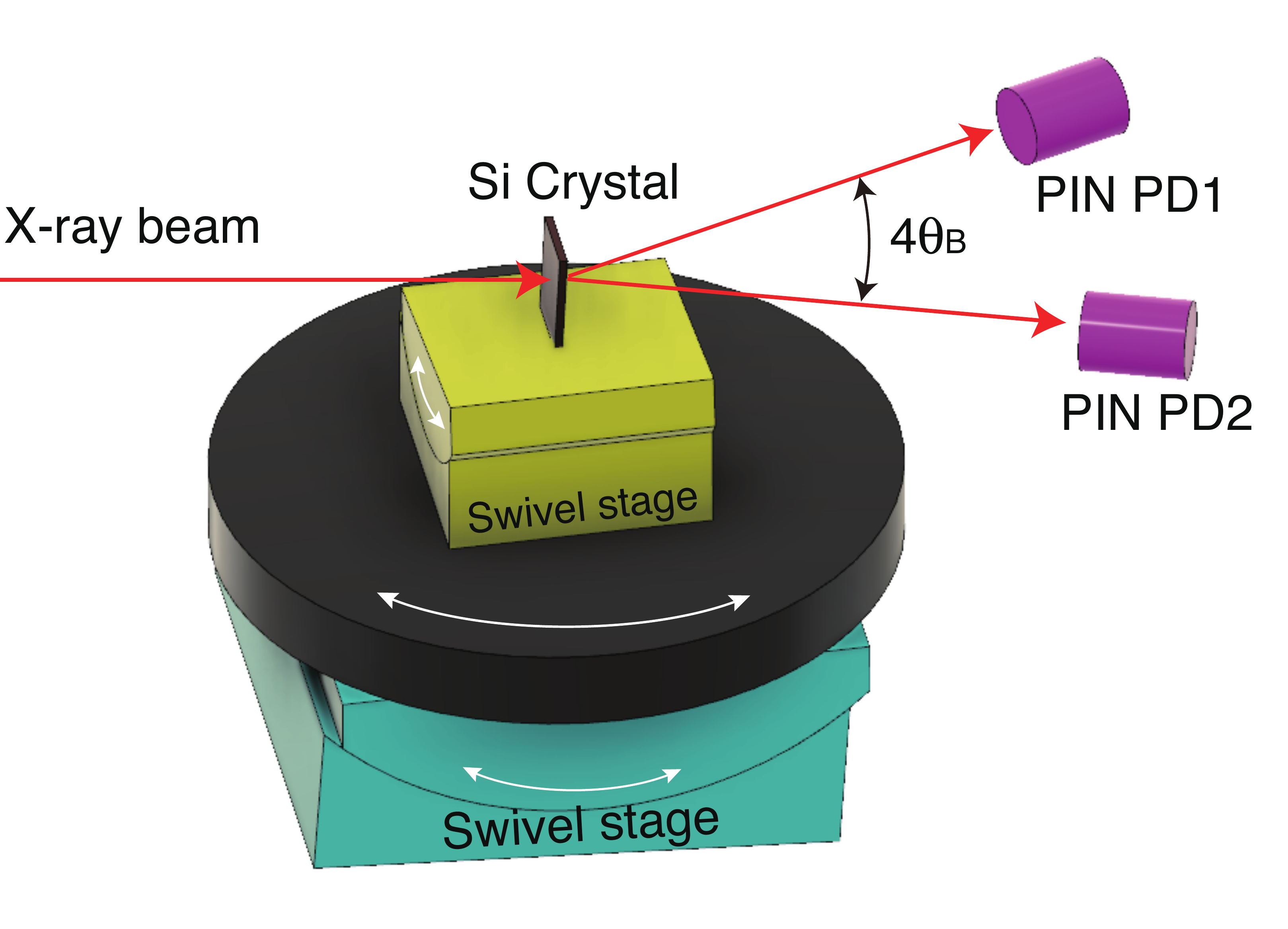}
\caption{Absolute energy measurement setup.
The X-ray beam is diffracted by a Si single crystal. 
Two PIN photodiodes monitor the diffracted beams. 
The rotary table, shown as a black disk, adjusts the mutual angle between the Si crystal 
and the X-ray beam so that the diffraction condition is satisfied. 
The two swivel stages adjust the tilt angles between the X-ray beam, 
the reciprocal lattice vector of the crystal and the rotation axis of the rotary table.
}
\label{fig:BondMethod}
\end{center}
\end{figure}

Our absolute X-ray energy monitor utilizes a method developed by Bond~\cite{Bond1960}.
As shown in Fig.~\ref{fig:BondMethod}, the monitor consists of a Si crystal plate and 
two X-ray sensors (PIN photo-diode); to control its directions, 
the crystal is mounted on a swivel stage, which is in turn placed on a rotary table and another swivel stage.  
The monitor locates two Bragg diffraction peaks, formed left and right of the beam, by rotating the table.
The rotation angle of the crystal between these two peaks is measured with a rotary encoder attached to the table. 
The incident X-ray wavelength $\lambda$ is derived using the relation
\begin{eqnarray}
   \lambda=2\,d(T,P)\sin \theta_{\rm B} \cos \theta_{\rm s1} \cos \theta_{\rm s2},
   \label{eq:Bragg eq}
\end{eqnarray}
where $d(T,P)$ is the smallest spacing between the crystal lattice planes at temperature $T$ and pressure $P$,
and $\theta_{\rm s1(s2)}$ are the deviations from the right angle defined by 
the incident beam and crystal's rotation axis (crystal's reciprocal lattice vector) directions.
Note that $\theta_{\rm s1(s2)}$ can be controlled by the swivel stages.
The crystal used in the present experiment is cut from the ingot of a standard reference crystal 
whose (220) crystal spacing, $d_{220}$, is calibrated at $T=22.5\;^{\circ}$C and $P=0$ atm (vacuum)~\cite{Cavagnero2004,Cavagnero2004E}; 
we conservatively quote $d_{220}$ as $192.01559\pm0.00002$ pm, considering inhomogeneity of the crystal\cite{Fujimoto2011}. 
The present monitor employs the (440) plane; thus $d_{440}=d_{220}/2$ is used in Eq.~(\ref{eq:Bragg eq}) with appropriate 
corrections for $T$ and $P$~\cite{Schoedel2001,Lyon1980,Hall1967}.

The heart of the monitor is the high precision, self-calibrating rotary encoder (called SelfA)~\cite{Watanabe2014}.
It has 12 optical sensors which ``read" gratings grooved along the circumference of a rotating disc attached to the table.
The encoder generates its response function, the relation between the true angle and the rotary encoder readings, using self-acquired data;
namely, it analyses all the outputs from all sensors and gratings based using the Fourier analysis method, 
and corrects non-linearity caused by, for example, 
eccentricity/inclination of the grating disc from the true rotation axis, 
and non-uniformity of the grating intervals.
One salient feature of the monitor is that the 12 sensors are distributed at every 1/3, 1/4, and 1/7 of the disc periphery
to maximize the order of Fourier components.

The swivel angles are adjusted in-situ by adjusting for the setting which yields the largest $\lambda$ in Eq.~(\ref{eq:Bragg eq}); 
thus the procedure ensures $\theta_{\rm s1(s2)}\simeq 0$.    

In this monitor system, the largest uncertainty arises from the angle determination by the rotary encoder: 
this is found to be 0.044\,arcsec which translates to a fractional uncertainty of 0.67\,ppm in $\lambda$.
The second-largest uncertainty stems from $d_{220}$, which is 0.1\,ppm.
All other uncertainties, such as corrections due to $T$ or $P$, non-zero $\theta_{\rm s1(s2)}$ are found to be negligible ($\ll 0.1$\,ppm).
The combined uncertainty is 0.67\,ppm or 0.02\,eV in the $\sim 29$\,keV beam.

\begin{figure}[t]
\begin{center}
\includegraphics[width=9cm,bb=0 0 550 350, clip]{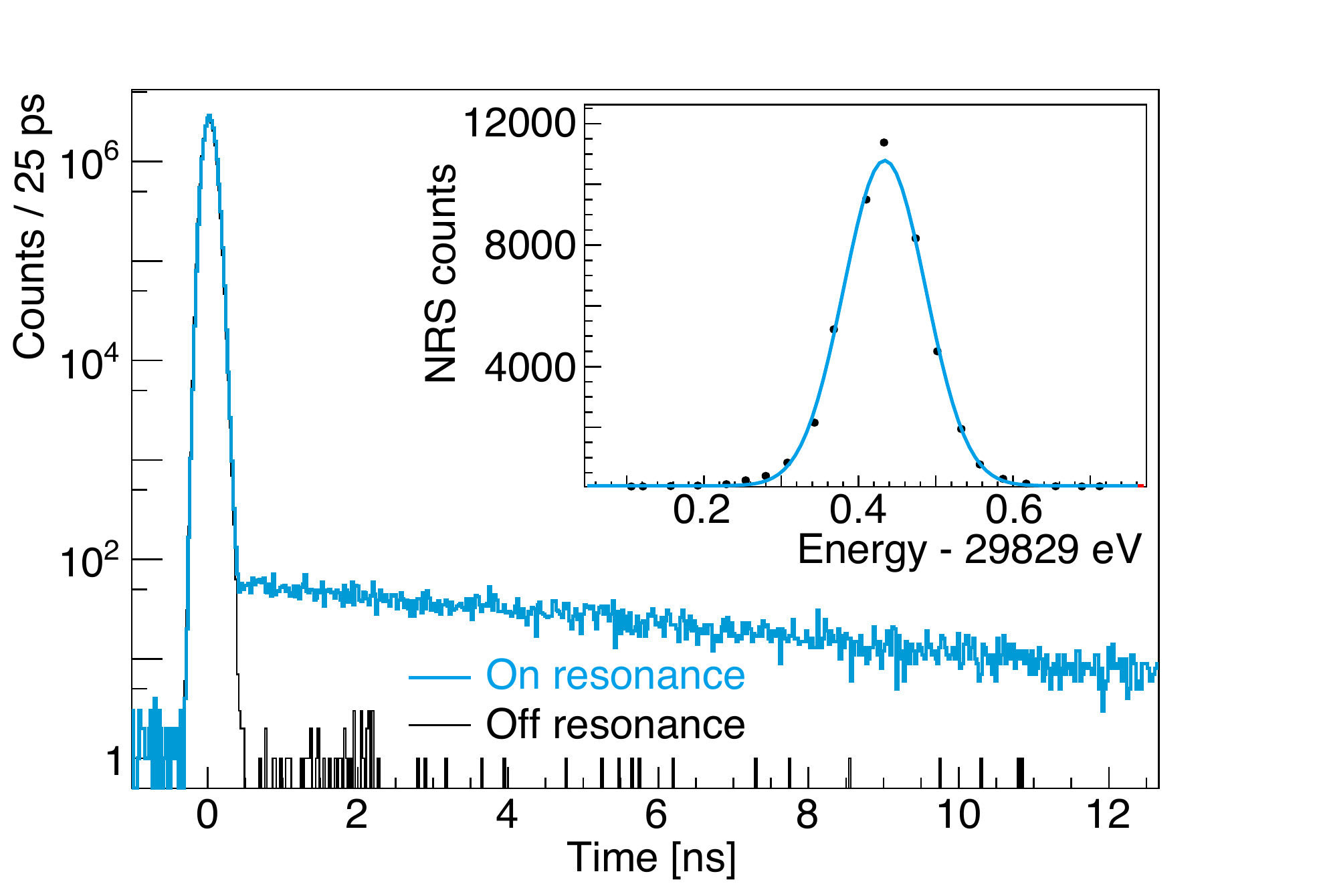}
\caption{$^{40}$K NRS spectrum. 
(lower plot) Examples of the temporal profiles at the on-resonance (blue histogram) or off-resonance (black histogram) incident X-ray energy.
(inset) The resonance curve with a Gaussian fit result (blue curve).}
\label{fig:K40}
\end{center}
\end{figure}

As an overall check, stability or reproducibility of the energy measurement 
is tested by monitoring the $^{40}$K NRS resonance energy~\cite{Seto2000}. 
A typical temporal profile from a $^{40}$K target is shown in Fig.~\ref{fig:K40}, together with 
the resonance curve in the inset. 
Actually the resonance energy was monitored 15 times 
in Run 1 and 2 with various mutual angles between the Si crystal and the rotary table.
The largest deviation from the average is found to be 0.07\,eV.
The quoted error in our absolute energy measurement is concluded from this deviation.

\section{Data and evaluation procedure\label{appendix:DataAnalysis}}

\subsection{Data taking procedure\label{appendix:DAT}}
The data presented in this article were taken in two separate beam times, 
each lasting about one week, in July (Run 1) and November (Run 2) 2018. 
During the resonance search in July, an energy range between 29189.6\,eV and 29198.0\,eV was scanned using the Si(440) monochromator.
A typical data taking procedure during the energy scan is as follows.
First, for each energy setting, the absolute value is measured with the absolute energy monitor, which takes about 180 seconds.
Then, data is accumulated for 1800 seconds (one standard run).
Finally, the energy is changed by a step of 0.08--0.12\,eV, corresponding to less than half of the Si(440) bandwidth.

After finding the resonance, Si(440) was replaced with Si(660) to take advantage of a better energy resolution 
(see Table~\ref{tab:beamsummary}). 
The data taking procedure using Si(660) is similar to that in Si(440) runs; however, 
the beam energy is varied in the immediate vicinity of the resonance, 
and the acquisition time is increased to 3600 seconds 
to compensate for the decrease in signal events (due to reduced beam intensity) within one run.

\subsection{Derivation of physics quantities\label{appendix:PHY}}
\subsubsection{Radiative width $\Gamma_{\gamma}^{\rm cr}$.}
To obtain the cross-band radiative width $\Gamma_{\gamma}^{\rm cr}$, 
the counting rates of the photoelectric absorption process $Y_{\rm pe}$, and the NRS process $Y_{\rm nrs}$ are compared. 
As already mentioned, the prompt peak consists predominantly of events caused by the photoelectric process 
(see below for the extraction procedure of $Y_{\rm pe}$ from the prompt peak). 
Its rate $Y_{\rm pe}$ is a product of three factors; the photoelectric absorption rate, 
the probability of yielding detectable X-ray photons, and various detection efficiencies.
Specifically, it is expressed by 
\begin{eqnarray}
   Y_{\rm pe}=\sigma_{\rm pe} \Phi_0 N_{\rm T} \langle \eta_{\rm pe}\epsilon_{\rm apd} \rangle \frac{\Delta \Omega}{4\pi},
   \label{eq:prompt rate}
\end{eqnarray}
where $\sigma_{\rm pe}$ represents the photoelectric total cross section, 
$\Phi_0$ the incident beam intensity (the number of X-rays per unit time), and
$N_{\rm T}$ the column density of $^{229}$Th nuclei in the target. 
The factor $\eta_{\rm pe}$ represents the probability that the photoelectric process produces X-ray fluorescences 
with energies inside our energy window (12-18\,keV). 
The last two factors are efficiencies of the detectors: the APD efficiency $\epsilon_{\rm apd}$ and 
the geometrical acceptance $\Delta \Omega/(4\pi)$.
The notation $\langle \eta_{\rm pe}\epsilon_{\rm apd} \rangle$ needs clarification:
it indicates that, since the APD efficiency depends on the X-ray energies, 
the energy average of $\eta_{\rm pe}$ is taken with a weight of $\epsilon_{\rm apd}$.
  
Similarly, the NRS rate is given by
\begin{eqnarray}
   Y_{\rm nrs}=\int dE \left\{ \sigma_{\rm nrs}(E) \frac{d\Phi}{dE}\right\} N_{\rm T} \langle \eta_{\rm ic}\epsilon_{\rm apd} \rangle
      \frac{\Delta \Omega}{4\pi} \epsilon_{\rm tw},
   \label{eq:NRS rate}
\end{eqnarray}
where $\sigma_{\rm nrs}(E)$ is the NRS cross section (given below) at the incident beam energy $E$, 
$d\Phi/dE$ the beam intensity per unit energy, 
$\eta_{\rm ic}$ the equivalent to $\eta_{\rm pe}$ for the NRS internal conversion process, 
and $\epsilon_{\rm tw}$ an efficiency due to the time window cut ($0.4-0.9$\,ns).
Note that, for the photoelectric process, the energy dependence of $\sigma_{\rm pe}$ can be ignored; 
for the NRS process, however, it is not possible to do so because $\sigma_{\rm nrs}$ is expected to have a 
much narrower width than that of the incident beam energy.  
The integration in Eq~.(\ref{eq:NRS rate}) is carried out as follows.
In the present experiment, $d\Phi/dE$ is well expressed by a Gaussian shape function
\begin{eqnarray}
 && \frac{d\Phi}{dE}=\Phi_0 f_{\rm b}(E), \\
 && f_{\rm b}(E)=\frac{1}{\sqrt{2\pi}\sigma_{\rm Xray}}
 \exp \left\{ -\frac{(E-E_0)^2}{2\sigma_{\rm Xray}^2} \right\},
 \label{eq:beam} 
\end{eqnarray}
where $E_0$ and $\sigma_{\rm Xray}$ are, respectively, the centre energy and the rms energy width of the beam.
The NRS cross section is given by the Breit-Wigner form~\cite{pdgBW}
\begin{eqnarray}
	\sigma_{\rm nrs}(E)= g_{\rm sp}\frac{\lambda_{\rm 2nd}^2}{\pi}
       \left[ \frac{\Gamma_{\gamma}^{\rm cr} \Gamma_{\rm t}/4 }{(E-E_{\rm 2nd})^2+(\Gamma_{\rm t}/2)^2} \right],
\end{eqnarray}
where $\lambda_{\rm 2nd}$ is the wavelength corresponding to $E_{\rm 2nd}$, 
$\Gamma_{\rm t} \;(\Gamma_{\gamma}^{\rm cr})$ the total (cross-band transition) width of the 29-keV level, 
and $g_{\rm sp}$ the spin multiplicity factor.
Note that the factor in the square brackets $[\cdots]$ may be replaced by 
$(\pi/2) \Gamma_{\gamma}^{\rm cr} \delta (E-E_{\rm 2nd})$ in the narrow-width limit, 
which is well justified in the present case.
The spin factor $g_{\rm sp}$ is given by 
\begin{eqnarray}
 && g_{\rm sp}=\frac{2I_{\rm e}+1}{2(2I_{\rm g}+1)}=\frac{1}{2}
 \label{eq:spin factor} 
\end{eqnarray}
where $I_{\rm e}(=5/2)$ and  $I_{\rm g}(=5/2)$ are the nuclear spin of the 29-keV and ground level, respectively, 
and 2 in the denominator is the spin multiplicity of the photon.
Convoluting $\sigma_{\rm nrs}$ with $f_{\rm b}(E)$, the first factor in Eq.~(\ref{eq:NRS rate}) turns out to be 
\begin{eqnarray}
	\int  \sigma_{\rm nrs}(E)\Phi_0\,f_{b}(E)\;dE
        =\frac{\lambda_{\rm 2nd}^2}{4} \frac{\Gamma_{\gamma}^{\rm cr}}{\sqrt{2\pi}\sigma_{\rm Xray}} \Phi_0
        \label{eq:production cross section 2nd}
\end{eqnarray}
when $E_0=E_{\rm 2nd}$.
In Eqs.~(\ref{eq:prompt rate}) and~(\ref{eq:NRS rate}), several factors are common, and thus the ratio becomes
\begin{eqnarray}
\frac{Y_{\rm nrs}}{Y_{\rm pe}}=\frac{\lambda_{\rm 2nd}^2}{\sigma_{\rm pe}}\;
\frac{\Gamma_{\gamma}^{\rm cr}}{4\sqrt{2\pi}\sigma_{\rm Xray}} 
\frac{\langle \eta_{\rm ic}\epsilon_{\rm apd} \rangle}{\langle \eta_{\rm pe}\epsilon_{\rm apd} \rangle} \epsilon_{\rm tw}.
\label{eq:ratio}
\end{eqnarray}
Among the quantities in Eq.~(\ref{eq:ratio}), $Y_{\rm nrs}/Y_{\rm pe}$, $\lambda_{\rm 2nd}$, $\sigma_{\rm Xray}$, and  $\epsilon_{\rm tw}$ 
can be determined from our own measurements, while $\sigma_{\rm pe}$ is tabulated~\cite{NIST-XCOM}. 
The ratio $\eta_{\rm ic}/\eta_{\rm pe}$ is expected to be $\sim 1$ since the processes are similar; 
this fact may be confirmed by theoretical estimates as well as the prompt and NRS experimental spectra. 
The detail is described in a separate paragraph below together with the procedure to extract $Y_{\rm pe}$ from the prompt peak.
The efficiency $\epsilon_{\rm tw}$ is essentially given 
by the integral of an exponential decay probability in the time range of 0.4--0.9\,ns; 
 in the actual calculation, however, a subtle effect is taken into account due to the time resolution of the APD detector.  
All relevant values and their errors are listed in Table \ref{table:radiative width}.
Inserting these values into Eq.~(\ref{eq:ratio}), 
$\Gamma_{\gamma}^{\rm cr}= 1.70 \pm 0.40\hspace{2mm}{\rm neV} $
is obtained, where the error is a quadrature sum of the individual errors listed in Table~\ref{table:radiative width}.
The main errors stem from $\epsilon_{\rm tw}$ which is sensitive to the half-life of the 29-keV level, 
and from $\sigma_{\rm Xray}$ which is determined by the Gaussian fit to the resonance curve.
The error next in size comes from $Y_{\rm nrs}/Y_{\rm pe}$, whose accuracy is determined by an error on $Y_{\rm nrs}$.
All the other errors are negligible compared to the three outlined above.

\begin{table}[t]
\caption{Numerical values used in the estimation of $\Gamma_{\gamma}^{\rm cr}$.}
\begin{center}
\begin{tabular}{lcc} \hline
Item    & value and error & unit    \\
\hline
$Y_{\rm nrs}/Y_{\rm pe}$ & $(1.84\pm 0.20)\times 10^{-7}$ & \\
$\lambda_{\rm 2nd}$ & $42.5\pm 0.00013$ & pm \\
$\sigma_{\rm Xray}$ & $0.041 \pm 0.006$ & eV   \\
$\epsilon_{\rm tw}$ & $0.039 \pm 0.006$  & \\
$\langle \eta_{\rm ic}\epsilon_{\rm apd} \rangle/\langle \eta_{\rm pe}\epsilon_{\rm apd} \rangle$ & $0.96 \pm 0.01$ &  \\
$\sigma_{\rm pe}$ & $15.4 \pm 0.8$ & kb \\
\hline 
\end{tabular}
\label{table:radiative width}
\end{center}
\end{table}

\begin{table*}[t]
\caption{Comparison of energy-averaged line strengths.}
\begin{center}
\begin{tabular}{cc|lll|ll} \hline
\multicolumn{2}{c}{Emission lines}& \multicolumn{3}{|c|}{Theoretical estimates} & \multicolumn{2}{c}{Experimental data} \\ 
Names & Energy & p.e. & i.c. M1 & i.c. E2 & Prompt & NRS \\ 
& [keV] & $\times 10^{-3}$ (\%) &  $\times 10^{-3}$ (\%)&  $\times 10^{-3}$ (\%) & (\%) & (\%) \\ \hline
$L_\iota$ & 11.1  & 0.26 (5.3) & 0.24 (5.0)& 0.24 (5.0) & ($0.6\pm 0.1$) & ($1.5\pm 0.6$)\\
$L_{\alpha_1},L_{\alpha_2}$ & 12.8-13.0 & 2.82 (56.9) & 2.58 (54.1)  & 2.57 (53.6) & ($46.6\pm 2.0$) & ($51.6\pm 4.9$) \\
$L_{\beta_{2,15}},L_{\beta_4},L_{\beta_6},L_\eta$ &14.5-15.6 & 0.55 (11.2) &  0.81 (17.0)&  0.42 (8.8) & ($11.9\pm 3.5$) & ($13.7\pm 5.3$)\\
$L_{\beta_1},L_{\beta_3},L_{\beta_5}$ & 16.2-16.4 & 1.11 (22.5) & 0.88 (18.6)& 1.34 (28.0) & ($37.2\pm 1.8$) & ($30.3\pm 4.5$)\\
$L_{\gamma_1},L_{\gamma_2},L_{\gamma_3}, L_{\gamma_6}$ & 19.0-19.6 & 0.22 (4.2) & 0.25 (5.3)&  0.22 (4.6)& ($3.7\pm 0.4$) & ($2.8\pm 2.0$)\\ \hline
Sum & 11.1-19.6 & 4.96 (100) & 4.76 (100)& 4.79 (100) & (100) & (100)\\ 
\hline
\end{tabular}
\label{List of line strengths}
\end{center}
\end{table*}
%

\subsubsection{The ratio $\langle \eta_{\rm ic}\epsilon_{\rm apd} \rangle /\langle \eta_{\rm pe}\epsilon_{\rm apd} \rangle$.}
Events originating from both photoelectric absorption and NRS scattering processes emit X-ray fluorescences in their relaxation processes. 
In the present experiment, 12-18 keV X-rays are detected; 
X-rays in this energy range are produced only when vacancies are created in the L shell.  
There are three subshells: L$_1$, L$_2$, and  L$_3$.
Each vacancy in the subshell emits characteristic X-rays, referred to as ``emission lines", 
by filling the vacancy with electrons in upper shells.
Once a subshell is specified, then the energy distribution of the emission lines and their strengths 
({\it i.e.} emission probability per one vacancy) are determined independent of its parent process. 
These data are compiled for all the $^{229}$Th subshells~\cite{IsotopeTable}.
Thus the determination of the X-ray emission spectra necessary to calculate 
$\langle \eta_{\rm ic}\epsilon_{\rm apd} \rangle /\langle \eta_{\rm pe}\epsilon_{\rm apd} \rangle$ boils down to the 
determination of the subshell distribution for each process.

For clarity, let's denote by $\eta_{\rm pe}^{i}$ or $\eta_{\rm ic}^{i}$ the probability of creating the L$_i$ subshell vacancy, 
and by $\eta^{ij}$ the strength of the line $j$ by the L$_i$ subshell vacancy 
($j$ runs over all possible emission lines usually labeled as $L_{\alpha_1}, L_{\alpha_2}, L_{\beta_1}, L_{\beta_{2,15}}, $ {\it etc}.).
For the photoelectric process, the values of $\eta_{\rm pe}^{i}$ are well established~\cite{IsotopeTable,Scofield1973}; they are 
(0.180, 0.257, 0.312, 0.749) for (L$_1$, L$_2$, L$_3$, $\sum L_i$), respectively.
On the other hand, it is somewhat more involved for the internal conversion process 
because two types of multipoles, M1 and E2, play a role in this case, and 
they yield different $\eta_{\rm ic}^{i}$ in general.
At this stage, it is convenient to introduce a quantity called the internal conversion coefficient, $\alpha_{\rm ic}$, 
defined by the ratio of the internal conversion width to the corresponding radiative width. 
The merits of introducing $\alpha_{\rm ic}$ are as follows.
First, calculation of $\alpha_{\rm ic}$ is expected to be reliable 
because the nuclear matrix element cancels out in the ratio.
Given $\alpha_{\rm ic}$ for a subshell L$_i$,  
$\eta_{\rm ic}^{i}$ is obtained by taking a ratio of $\alpha_{\rm ic}$ for the subshell to that for the total (all shells and subshells).
Another merit is that since $\eta_{\rm ic}^{i}$ in the present case is a linear mixture of the M1 and E2 components, 
it is bounded by two extremes, pure M1 and pure E2.
Actually $\eta_{\rm ic}^{i}$ is calculated, for these extremes, using the BrIcc v2.3S code~\cite{Bricc}(bricc.anu.edu.au/index.php).
The values of $\eta_{\rm ic}^{i}$ thus obtained are (0.673, 0.078, 0.005, 0.756) for M1, 
and (0.011, 0.363, 0.359, 0.733) for E2.

Table~\ref{List of line strengths} shows the list of energy-averaged line strengths 
defined as $\sum_{i=1}^3 \eta_{\rm pe,ic}^{i}\eta^{ij} \varepsilon_{\rm apd}(j)$,  
where $\varepsilon_{\rm apd}(j)$ is evaluated at the energy of $j$. 
In the table, the lines with similar energies are grouped together for convenience.
In the column of the theoretical estimates, the energy-averaged line strengths for the photoelectric process (p.e.), 
and the internal conversion  process of pure M1 (i.c. M1) and of pure E2 (i.c. E2) are compared.  
The values in the parentheses are those normalized to the sum, which is shown in the bottom row.
The energy dependence of $\varepsilon_{\rm apd}$, together with the APD response function, is studied 
by a separate experiment at KEK-PF~\cite{Masuda2019}; it is expressed by 
$\varepsilon_{\rm apd}(E)\simeq 1-\exp[-\sigma_{\rm Si}(E)\rho_{\rm Si} L_{\rm Si}]$, 
where $\sigma_{\rm Si}(E)$ is the photoelectric absorption cross section~\cite{NIST-XCOM} of Si at the energy $E$, 
$\rho_{\rm Si}$ the density and $L_{\rm Si}$ an effective thickness of the Si sensor.
Some uncertainties exit in determining $L_{\rm Si}$, but their effect is negligible for the ratio calculation. 
As seen in the bottom row of Table~\ref{List of line strengths}, the sums of the energy-averaged line strengths,
$\langle \eta_{\rm pe,ic} \varepsilon_{\rm apd}\rangle=\sum_j \sum_{i=1}^3 \eta_{\rm pe,ic}^{i}\eta^{ij} \varepsilon_{\rm apd}(j)$, 
are similar for all three cases.
Taking into account the fact that some lines (11.1 and 19.6\,keV) partially fall outside the energy window, 
the ratio $\langle \eta_{\rm ic}\epsilon_{\rm apd} \rangle /\langle \eta_{\rm pe}\epsilon_{\rm apd} \rangle$ is found to be 
$0.96 \pm 0.01$, where the error comes mainly from the ambiguity in the mixing ratio of M1 and E2 components.

\begin{figure*}[tb]
\begin{center}
\includegraphics[width=9cm,bb=0 0 600 615, clip]{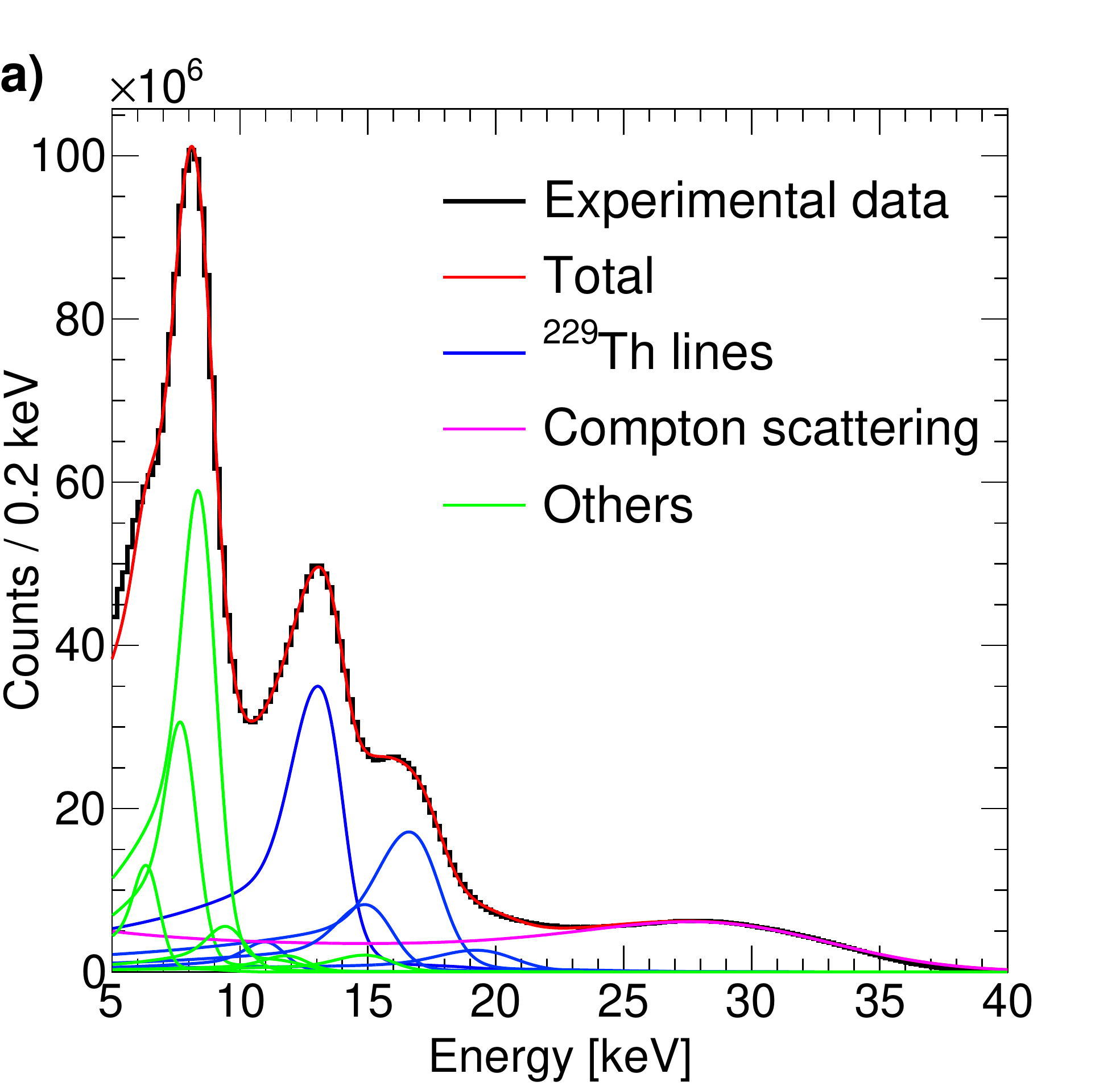}
\includegraphics[width=9cm,bb=0 0 600 615, clip]{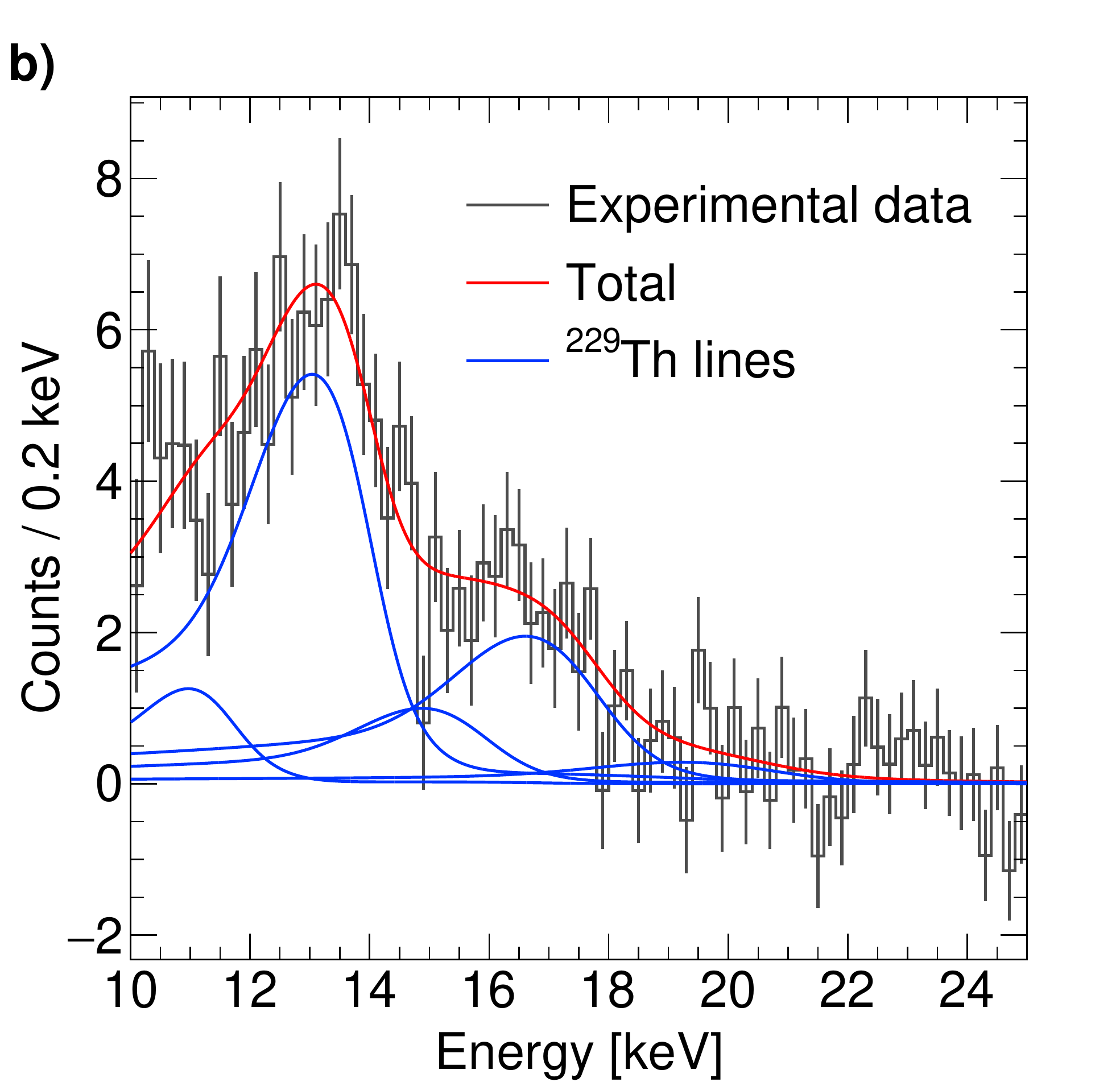}
\caption{Energy spectra of the prompt and NRS events. a) Prompt energy spectrum.
The coloured lines are various X-ray emission lines convoluted with the APD energy response function: 
the photoelectric lines listed in Table~\ref{List of line strengths} (in blue), 
the Compton scattering (in magenta), and the K$_{\alpha,\beta}$ lines of Cu, Zn, and Fe (in green).
The strengths of these lines are adjusted to give the best fit to the data. 
The sum, shown in red, well reproduces the data above 7\,keV.\hspace{1mm}
b) NRS energy spectrum, which is obtained by subtracting the 
off-resonance data from the on-resonance data. 
The coloured lines are the fit results of X-ray emission lines.
Note that there is no contribution from the Compton scattering or Cu/Zn/Fe lines. 
Both experimental data sets are normalized to a 3600\,s run.}
\label{fig:Energy-Spectrum-prompt-NRS}
\end{center}
\end{figure*}
%

In the present experiment, each energy-averaged line strength may be deduced from the measurements 
and may be compared with the theoretical estimates.
Figure~\ref{fig:Energy-Spectrum-prompt-NRS}a shows the energy spectrum of the prompt peak (black histogram).
The coloured lines are various X-ray emission lines convoluted with the APD energy response function.
Specifically, the blue lines are the photoelectric lines listed in Table~\ref{List of line strengths}, 
the magenta the Compton scattering, and the green lines are K$_{\alpha,\beta}$ lines of Cu, Zn, Fe, and others
 (originated from the brass collimator {etc.}).
The origin of these background contaminations is confirmed by separate measurements with a Silicon-Drift-Detector (SDD, R\"ontec Xflash) 
carried out during the experiment. Note that the SDD has much better energy resolution than our APD, although it is slow. 
The strengths of these lines are adjusted to give the best fit to the data. 
The sum of these lines, shown in red, well reproduces the data above 7\,keV.
Figure~\ref{fig:Energy-Spectrum-prompt-NRS}b shows the NRS energy spectrum; this is obtained by 
subtracting the non-resonant data from the resonant data. 
The coloured lines are the fit results of X-ray emission lines, 
obtained in a similar manner to that for the prompt spectrum.
The sum of the lines agrees well with the real data.
From the fits, relative strengths of the emission lines can be deduced; 
these are compared with the theoretical estimates in Table~\ref{List of line strengths} 
(see the values in parentheses).
Reasonable agreement is obtained between the fit results and theoretical estimates for both processes.

The fit results may also be used to extract the photoelectric rate from the prompt rate.
The fraction of the Compton and background lines (Cu/Zn/Fe {\it etc.}) underneath 
of the energy range of 12-18\,keV is found to be $0.097\pm 0.004$.
This value is used to obtain $Y_{\rm pe}/Y_{\rm nrs}$ in Table~\ref{table:radiative width}.

\subsubsection{Branching ratio.}
The branching ratio of the 29-keV level to the ground state by radiative transitions is 
defined by 
\begin{eqnarray}
 B_{\gamma}^{\rm cr}=\frac{\Gamma_{\gamma}^{\rm cr}}{\Gamma_{\gamma}^{\rm cr}+\Gamma_{\gamma}^{\rm in}},
 \label{eq:Br cr def} 
\end{eqnarray}
where $\Gamma_{\gamma}^{\rm in}$ is the inband transition width.
Among the two widths, $\Gamma_{\gamma}^{\rm cr}$ is already obtained (see Eq.~(\ref{eq:radiative width})). 
The inband transition width has been reported by two experimental groups~\cite{Barci2003 ,Kroger1976}, 
giving $\Gamma_{\gamma}^{\rm in}/\hbar=(2.161\pm 0.240)\times 10^{7}\,{\rm s}^{-1}$ and 
$(2.238\pm 0.531)\times 10^{7}\,{\rm s}^{-1}$.
Taking a weighed average of these ($\Gamma_{\gamma}^{\rm in}=14.3\pm 1.4$\,neV), $B_{\gamma}^{\rm cr}$ is found to be
\begin{eqnarray}
 B_{\gamma}^{\rm cr}=0.106 \pm 0.027
 \label{eq:brachning ratio of gamma cross} 
\end{eqnarray}
or equivalently $B_{\gamma}^{\rm cr}=1/(9.4 \pm 2.4)$.

\subsubsection{Isomer production rate.}
The isomer production rate is given by the product of the 29-keV production rate and 
the branching ratio to the isomer state.
In this case, the branching ratio should include not only the radiative transition but also the internal conversion processes.
Denoting as $B_{\gamma+{\rm ic}}^{\rm in}$, it is expressed by 
\begin{eqnarray}
 B_{\gamma+{\rm ic}}^{\rm in}=\frac{\Gamma_{\gamma}^{\rm in}+\Gamma_{\rm ic}^{\rm in}}{\Gamma_{\rm t}}=\frac{\Gamma_{\gamma}^{\rm in}(1+\alpha_{\rm ic}^{\rm in})}{\Gamma_{\rm t}},
\end{eqnarray}
where $\Gamma_{\gamma}^{\rm in}$ and $\Gamma_{\rm ic}^{\rm in}$ represents, respectively, the radiative and the internal conversion 
width of the inband transition of the 29-keV level. 
On the right-hand side, the ratio $\alpha_{\rm ic}^{\rm in}=\Gamma_{\rm ic}^{\rm in}/\Gamma_{\gamma}^{\rm in}$ is introduced.
There is a theoretical estimate for $\alpha_{\rm ic}^{\rm in}$: $\alpha_{\rm ic}^{\rm in}\simeq 225$~\cite{NSDF}.
For the denominator, the experimental result of our half-life measurement is used: 
$\Gamma_{\rm t}=(\hbar \ln 2)/T_{1/2}=5550\pm 270\,{\rm neV}$. 
Putting these values together, $B_{\gamma+{\rm ic}}^{\rm in}$ is found to be $B_{\gamma+{\rm ic}}^{\rm in}=0.58 \pm 0.07$.
The 29-keV production rate is given by Eq.~(\ref{eq:production cross section 2nd}), where 
approximate values of $\Phi_0$ and $N_{\rm T}$ can be inserted.
The product of these two factors is $\sim 2.5 \times 10^4$\,s$^{-1}$.
As a byproduct, it is possible to extract $\alpha_{\rm ic}^{\rm cr}=\Gamma_{\rm ic}^{\rm cr}/\Gamma_{\gamma}^{\rm cr}$.
Rewriting $\alpha_{\rm ic}^{\rm cr}$ as
\begin{eqnarray}
 && \alpha_{\rm ic}^{\rm cr}+1=\frac{\Gamma_{\rm ic}^{\rm cr}+\Gamma_{\gamma}^{\rm cr}}{\Gamma_{\gamma}^{\rm cr}}=
 \frac{\Gamma_{\rm t}-\Gamma_{\gamma}^{\rm in}(1+\alpha_{\rm ic}^{\rm in})}{\Gamma_{\gamma}^{\rm cr}}
 \label{eq:alpha cr} 
\end{eqnarray}
and inserting into the right-hand side the values already obtained, 
$\alpha_{\rm ic}^{\rm cr}$ is found to be $\alpha_{\rm ic}^{\rm cr}=1370 \pm 410$.

\end{appendix}

\vspace{1cm}

\end{document}